\documentclass[12pt]{article}

\newcommand{\be}{\begin{equation}}
\newcommand{\ee}{\end{equation}}
\newcommand{\bi}[1]{\vspace{-3mm} \bibitem{#1}}

\usepackage{epsfig,amsmath,amssymb,graphics,graphicx} 

\begin{document}
%%%%%%%%%%%%%%%%%%%%%%%%%%%%%%%%%%%%%%%%%%%%%%%%%%%%%%%%%%%%%%%%%
\begin{center}

{\it Chaos, Solitons and Fractals. Vol.67. (2014) 26-37.}
\vskip 3mm

{\bf \large Flow of Fractal Fluid in Pipes: \\
Non-Integer Dimensional Space Approach} \\

\vskip 7mm
{\bf \large Vasily E. Tarasov} \\
\vskip 3mm

{\it Skobeltsyn Institute of Nuclear Physics,\\ 
Lomonosov Moscow State University, Moscow 119991, Russia} \\
{E-mail: tarasov@theory.sinp.msu.ru} \\

\begin{abstract}
Using a generalization of vector calculus for the case 
of non-integer dimensional space 
we consider a Poiseuille flow of 
an incompressible viscous fractal fluid in the pipe. 
Fractal fluid is described as a continuum 
in non-integer dimensional space. 
A generalization of the Navier-Stokes equations 
for non-integer dimensional space, 
its solution for steady flow of 
fractal fluid in a pipe and corresponding 
fractal fluid discharge are suggested.
\end{abstract}

\end{center}
%%%%%%%%%%%%%%%%%%%%%%%%%%%%%%%%%%%%%%%%%%%%%%%%%%%%%%%%%
\section{Introduction}

A cornerstone of fractal fluids is 
the non-integer dimension \cite{Fractal1,Fractal2,CM}. 
The mass of fractal fluid  
satisfies a power law relation $M \sim R^{D}$, 
where $M$ is the mass of the ball region with radius $R$,
and $D$ is the mass dimension \cite{TarasovSpringer}. 
Fractal fluid can be described by four different approaches:
(1) Using the methods of "Analysis on fractals" 
\cite{Kugami,Strichartz-1,Strichartz-2,Harrison,Kumagai,DGV} 
it is possible to describe fractal media;
(2) An application of fractional-differential continuum models 
suggested in \cite{CCC2001,CCC2002}, and then developed in 
\cite{CCC2004a,CCC2004c,CCSPZ2009,CCC2009,Yang2013a,Yang2013b},
where so-called local fractional derivatives \cite{Yang2013c} 
are used; 
(3) Applying fractional-integral continuum models suggested in 
\cite{PLA2005-1,AP2005-2,IJMPB2005-2,MPLB2005-1,TarasovSpringer} 
(see also \cite{MOS-3}-\cite{MOS-5} and \cite{Bal-1}-\cite{Bal-4}), where
integrations of non-integer orders and 
a notion of density of states \cite{TarasovSpringer} are used;  
(4) Fractal media can be described by using
the theory of integration and differentiation for
a non-integer dimensional space \cite{Collins,Stillinger,PS2004}
(see also \cite{CNSNS2015,JMP2014}.

%%%%%%%%%%%%%%%%%%%%%%%%%%%%%%%%%%%%%%%%%%%%%%%%%%%%%%%%%%%

Let us note that main difference of 
the continuum models with non-integer dimensional spaces
form the fractional continuum models suggested in 
\cite{PLA2005-1,AP2005-2,IJMPB2005-2,MPLB2005-1,TarasovSpringer}
may be reduced to the following.
(a) Arbitrariness in the choice of the numerical factor
in the density of states is fixed by the equation
of the volume of non-integer dimensional ball region.
(b) In the fractional continuum models suggested 
in \cite{PLA2005-1,AP2005-2,TarasovSpringer}, 
the differentiations are integer orders whereas
the integrations are non-integer orders.
In the continuum models with non-integer dimensional spaces
the integrations and differentiations are defined 
for the spaces with non-integer dimensions.

In this paper, we consider approach based on
the non-integer dimensional space.
The power law $M \sim R^{D}$ can be naturally 
derived by using the integrations 
in non-integer dimensional space \cite{Collins},
where the mass dimension of fractal fluid
is connected with the dimension of this space.
A vector calculus for non-integer dimensional space 
proposed in this paper
allows us to use continuum models with
non-integer dimensional spaces to describe
for fractal fluids.
This is due to the fact that
although the non-integer dimension  
does not reflect completely the geometric
properties of the fractal media, it nevertheless permits
a number of important conclusions about the behavior
of fractal structures.
Therefore continuum models with non-integer dimensional spaces
can be successfully used to describe fractal fluids.

Integration over non-integer dimensional spaces 
are actively used in the theory of critical phenomena and 
phase transitions in statistical physics 
\cite{WilsonFisher,WK1974}, and 
in the dimensional regularization of
ultraviolet divergences 
in quantum field theory \cite{HV1972,Leibbrandt,Collins}. 
The axioms for integrations in non-integer dimensional space are proposed in \cite{Wilson,Stillinger} and this type of integration is considered in the book by Collins \cite{Collins}
for rotationally covariant functions. 
In the paper \cite{Stillinger} a mathematical basis of integration on non-integer dimensional space is given. 
Stillinger \cite{Stillinger} suggested a generalization of 
the Laplace operator for non-integer dimensional spaces also. 
Using a product measure approach,  
the Stillinger's methods \cite{Stillinger}
has been generalized by Palmer and Stavrinou \cite{PS2004} 
for multiple variables case
with different degrees of confinement in orthogonal directions. 
The scalar Laplace operators suggested 
by Stillinger in \cite{Stillinger} 
and Palmer, Stavrinou in \cite{PS2004}
for non-integer dimensional spaces, 
have successfully been used for 
effective descriptions in physics and mechanics. 
The Stillinger's form of Laplacian 
for the Schr\"odinger equation in non-integer dimensional space 
is used by He \cite{XFHe1,XFHe2,XFHe3} to describe
a measure of the anisotropy and 
confinement by the effective non-integer dimensions. 
Quantum mechanical models with non-integer (fractional) 
dimensional space have been discussed in 
\cite{Stillinger,PS2004,Thilagam1997b,MA2001a,MA2001b,QM1,QM5}
and \cite{Muslih2010,MA2012,QM8,QM9}.
Recent progress in non-integer dimensional space approach
also includes description of
the fractional diffusion processes in
non-integer dimensional space in \cite{LSTLRL}, and
the electromagnetic fields in non-integer dimensional space
in \cite{MB2007,BGG2010,MSBR2010}
and \cite{ZMN2010,ZMN2011a,ZMN2011b,ZMN2011c,ZMN}. 

%%%%%%%%%%%%%%%%%%%%%%%%%%%%%%%%%%%%%%%%%%%%%%%%%%%%%

Unfortunately, \cite{Stillinger,PS2004} proposed 
only the second order differential operators 
for scalar fields in the form of the scalar Laplacian 
in the non-integer dimensional space. 
A generalization of the vector Laplacian \cite{VLap}
for the non-integer dimensional space is not suggested
in \cite{Stillinger,PS2004}. 
The first order operators such as gradient, 
divergence, curl operators  
are not considered in \cite{Stillinger,PS2004} also.
In the work \cite{ZMN} 
the gradient, divergence, and curl operators
are suggested only as approximations of the square of 
the Laplace operator.
Consideration only the scalar Laplacian 
in the non-integer dimensional space approach
greatly restricts us in application of continuum models
with non-integer dimensional spaces for fractal fluids and material.
For example, we cannot use the Stillinger's form of Laplacian
for vector field ${\bf v}({\bf r},t)$ in hydrodynamics of fractal fluids, in fractal theory of elasticity and thermoelasticity, 
in electromagnetic theory of fractal media
to describe processes
in the framework non-integer dimensional space approach.

In this paper, we propose to use a vector calculus
for non-integer dimensional space, and we define
the first and second orders differential vector
operations such as gradient, divergence, 
the scalar and vector Laplace operators 
for non-integer dimensional space.
In order to derive the vector differential operators 
in non-integer dimensional space
we use the method of analytic continuation in dimension.
For simplification we consider rotationally covariant 
scalar and vector functions that are independent of angles.
It allows us to reduce differential equations 
in non-integer dimensional space to
ordinary differential equations with respect to $r$. 
The proposed operators allow us to describe fractal media
to describe processes in the framework of continuum models
with non-integer dimensional spaces. 
In this paper we describe a Poiseuille flow of 
an incompressible viscous fractal fluid in the pipe. 
A generalization of the Navier-Stokes equation 
for non-integer dimensional space to describe 
for fractal fluid are suggested. 
A solution of this equation for steady flow of 
fractal fluid in a pipe and corresponding 
fractal fluid discharge are derived.

%%%%%%%%%%%%%%%%%%%%%%%%%%%%%%%%%%%%%%%%%%%%%%%%%%%%%%%%%
\section{Fractal fluids}

A basic characteristic of fractal fluids
is the non-integer dimensions such as mass 
or "particle" dimensions \cite{TarasovSpringer}. 
For fractal fluids 
the number of particles $N_D(W)$ or mass $M_D(W)$ 
in any region $W \subset \mathbb{R}^3$ of this fluid 
increase more slowly 
than the 3-dimensional volume $V_3(W)$ of this region.
For the ball region $W$ with radius $R$ in 
an isotropic fractal fluid, 
this property can be described by the relation
between the number of particles $N_D(W)$
in the region $W$ of fractal fluid, and 
the radius $R$ in the form
\be \label{NDW} 
N_D(W) = N_0 (R/ R_0)^D , \quad  R/R_0 \gg 1 , 
\ee
where $R_0$ is the characteristic size of fractal fluid
such as a minimal scale of self-similarity 
of a considered fractal fluid. 
The number $D$ is called the "particle" dimension. 
It is a measure of how the fluid particles
fill the space.
The parameter $D$ does not depend on the shape 
of the region $W$.
Therefore fractal fluids can be considered 
as fluid with non-integer "particle" or mass dimension. 

If the fractal fluid consists of particles 
with identical masses $m_0$, then relation (\ref{NDW}) gives
\be \label{MDW} 
M_D(W) = M_0 (R/ R_0)^D , \quad  R/R_0 \gg 1 , 
\ee
where $M_0=m_0 \, N_0$. 
In this case, the mass dimension coincides
with the "particle" dimension.

As the basic mathematical tool for
continuum models of fractal fluids,
we propose to use the integration and differentiation
in non-integer dimensional spaces.
In Section 7, we will show that 
the power-law relation (\ref{MDW}) for 
an isotropic fractal fluid  
can be naturally derived by using the integration over 
non-integer dimensional space, where
the space dimension is equal to
the mass dimension of fractal fluid. 

%%%%%%%%%%%%%%%%%%%%%%%%%%%%%%%%%%%%%%%%%%%%%%%%%%%%%%%%%

In order to describe fractal fluid by 
continuum models with non-integer dimensional spaces,
we use the concepts of 
density of states $c_3(D,{\bf r})$
that describes how closely packed permitted 
places (states) in the space $\mathbb{R}^3$, 
where the fractal fluid is distributed.
The expression $dV_D({\bf r})=c_3(D,{\bf r})dV_3$ 
is equal to the number of permitted places (states) 
between $V_3$ and $V_3 +dV_3$ in $\mathbb{R}^3$. 
The notation $d^D {\bf r}$ 
also will be used instead of $ d V_D ({\bf r})$.
Note that density of states and 
distribution function are different concepts, 
and it is impossible to describe 
all properties of fractal fluids
by the distribution function only.

For fractal fluids, we can use the equation
\be 
dN_D(W)= n({\bf r}) \, d V_D ({\bf r}) , 
\ee
where $n({\bf r})$ is a concentration of particles
that describes a distribution of number of particles
on a set of permitted places (possible states).
The density of states is chosen such that
$d V_D ({\bf r}) = c_3(D,{\bf r}) \, dV_3$
describes the number of permitted states in $dV_3$.

The form of the function $c_3(D,{\bf r})$ 
is defined by symmetries of 
considered problem and properties of 
the described fractal fluid.
A general property of density of states
for fractal fluids is a power-law type 
of these functions that reflects a scaling property 
(fractality) of the fractal fluid. 
To simplify our consideration in this paper 
we will consider only isotropic fractal fluids 
with density of states that is independent of angles.
In this case, the form of density of states is
defined such that $dV_D$ is an elementary volume
of the non-integer dimensional space.

In the continuum models of fractal fluids,
we should work with the dimensionless variables
$x/R_0 \to x$, $y/R_0 \to x$, $z/R_0 \to x$, 
${\bf r}/R_0 \to {\bf r}$,
in order to physical quantities of 
fractal fluids have correct physical dimensions.

%%%%%%%%%%%%%%%%%%%%%%%%%%%%%%%%%%%%%%%%%%%%%%%%%%%%%%%%%
\section{Vector differential operators 
in non-integer dimensional space}

To derive equations for vector differential operators 
in non-integer dimensional space,
we use equations for the differential operators 
in the spherical (and cylindrical) coordinates in $\mathbb{R}^n$ 
for arbitrary $n$ to highlight the explicit 
relations with dimension $n$.
Then the vector differential operators  
for non-integer dimension $D$ 
can be defined by continuation in 
dimension from integer $n$ to non-integer $D$. 
To simplify we will consider only scalar 
fields $\varphi$ and vector fields ${\bf v}$ 
that are independent of angles 
\[ \varphi({\bf r}) =\varphi(r) ,
\quad {\bf v}({\bf r})={\bf v} (r) = v_r\, {\bf e}_r , \]
where $r=|{\bf r}|$ is the radial distance,
${\bf e}_r={\bf r} /r$ is the local orthogonal unit 
vector in the directions of increasing $r$,  
and $v_r=v_r(r)$ is the radial component of ${\bf v}$.
We will work with rotationally covariant functions only. 
This simplification is analogous to the simplification
for definition of integration over non-integer dimensional space
described in Section 4 of the book \cite{Collins}.

%%%%%%%%%%%%%%%%%%%%%%%%%%%%%%%%%%%%%%%%%%%%%%%%%%%%%%%%%%%
\subsection{Vector differential operators 
for spherical and cylindrical cases}

%%%%%%%%%%%%%%%%%%%%%%%%%%%%%%%%%%%%%%%%%%%%%%%%%%%%%%%%%%%
%%%     D DIMENSION

Using the continuation from integer $n$ 
to arbitrary non-integer $D$,  
we can get explicit definitions of  differential operators
for non-integer dimensional space in the following forms.
Note that the same expressions can be obtained by using 
the integration in non-integer dimensional space
and the correspondent Gauss's theorem \cite{CNSNS2015}

Let us define the differential vector operations
such as  gradient, divergence, the scalar and 
vector Laplacian for non-integer dimensional space. 
For simplifications, we assume that
the vector field ${\bf v}={\bf v}({\bf r})$ 
be radially directed and the scalar and vector 
fields $\varphi({\bf r})$, ${\bf v}({\bf r})$ 
are not dependent on the angles.

The divergence in non-integer dimensional space 
for the vector field ${\bf v}={\bf v}(r)$ is
\be \label{Div-D}
\operatorname{Div}^{D}_{r} {\bf v} = 
\frac{\partial v_r}{\partial r} + \frac{D-1}{r} \, v_r.
\ee

The gradient in non-integer dimensional space 
for the scalar field $\varphi=\varphi (r)$ is
\be \label{Grad-D}
\operatorname{Grad}^{D}_r \varphi = \frac{\partial \varphi}{\partial r} \, {\bf e}_r .
\ee

The scalar Laplacian in non-integer dimensional space 
for the scalar field $\varphi=\varphi (r)$ is
\be \label{S-Delta-D}
^S\Delta^{D}_r \varphi= 
\operatorname{Div}^{D}_r \operatorname{Grad}^{D}_{r} \varphi =
\frac{\partial^2 \varphi}{\partial r^2} + \frac{D-1}{r} \, 
\frac{\partial \varphi}{\partial r} .
\ee

The vector Laplacian in non-integer dimensional space 
for the vector field ${\bf v}=v(r) \, {\bf e}_r$ is
\be \label{V-Delta-D}
^V\Delta^{D}_r {\bf v} = 
\operatorname{Grad}^{D}_r \operatorname{Div}^{D}_{r} {\bf v} =
\Bigl(
\frac{\partial^2 v_r}{\partial r^2} + \frac{D-1}{r} \, 
\frac{\partial v_r}{\partial r}  -  \frac{D-1}{r^2} \, v_r
\Bigr) \, {\bf e}_r.
\ee

If $D=n$, equations (\ref{Div-D}-\ref{V-Delta-D})
give the well-known formulas for
integer dimensional space $\mathbb{R}^n$.

%%%%%%%%%%%%%%%%%%%%%%%%%%%%%%%%%%%%%%%%%%%%%%%%%%%%%%%%%%%
%%%     D DIMENSION  - CYLINDER

Let us consider a case of axial symmetry 
of the fluid, where the fields 
$\varphi(r)$ and ${\bf v}(r)=v_r(r) \, {\bf e}_r$
are also axially symmetric.
We will direct the $Z$-axis along the axis of symmetry. 
Therefore we use a cylindrical coordinate system.

The divergence in non-integer dimensional space 
for the vector field ${\bf v}={\bf v}(r)$ is
\be \label{Div-DC}
\operatorname{Div}^{D}_{r} {\bf v} = 
\frac{\partial v_r}{\partial r} + \frac{D-2}{r} \, v_r.
\ee

The gradient in non-integer dimensional space 
for the scalar field $\varphi=\varphi (r)$ is
\be \label{Grad-DC}
\operatorname{Grad}^{D}_r \varphi = 
\frac{\partial \varphi}{\partial r} \, {\bf e}_r .
\ee

The scalar Laplacian in non-integer dimensional space 
for the scalar field $\varphi=\varphi (r)$ is
\be \label{S-Delta-DC}
^S\Delta^{D}_r \varphi = 
\frac{\partial^2 \varphi}{\partial r^2} + \frac{D-2}{r} \, 
\frac{\partial \varphi}{\partial r} .
\ee

The vector Laplacian in non-integer dimensional space 
for the vector field ${\bf v}=v(r) \, {\bf e}_r$ is
\be \label{V-Delta-DC}
^V\Delta^{D}_r {\bf v} = 
%%%\operatorname{Grad}^{D}_r \operatorname{Div}^{D}_{r} {\bf v} =
\Bigl(
\frac{\partial^2 v_r}{\partial r^2} + \frac{D-2}{r} \, 
\frac{\partial v_r}{\partial r}  -  \frac{D-2}{r^2} \, v_r
\Bigr) \, {\bf e}_r .
\ee

%%%%%%%%%%%%%%%%%%%%%%%%

Equations (\ref{Div-DC}-\ref{V-Delta-DC}) can be
easy generalized for the case $\varphi=\varphi(r,z)$ and 
${\bf v}(r,z)=v_r(r,z) \, {\bf e}_r+ v_r(r,z) \, {\bf e}_z$.
In this case the curl operator for ${\bf v}(r,z)$
is different from zero, and
\be \label{Curl-DC}
\operatorname{Curl}^{D}_r {\bf v} = 
\left( \frac{\partial v_r}{\partial z} -
\frac{\partial v_z}{\partial r} \right) \, {\bf e}_{\theta} .
\ee

%%%%%%%%%%%%%%

For $D=3$ equations (\ref{Div-D}) - (\ref{V-Delta-DC}) and 
(\ref{Curl-DC}) give the well-known
expressions for the gradient, divergence, curl operator,
scalar and vector Laplacian operators

The suggested operators for $0<D<3$ allow us to reduce 
$D$-dimensional vector differentiations 
(\ref{Div-D}) - (\ref{V-Delta-D}) and 
(\ref{Div-DC}) - (\ref{V-Delta-DC})
to derivatives with respect to $r=|{\bf r}|$. 
It allows us to reduce partial differential equations for
fields in non-integer dimensional space to
ordinary differential equations with respect to $r$.

%%%%%%%%%%%%%%%%%%%%%%%%%%%%%%%%%%%%%%%%%%%%%%%%%%%%%%%%%
\subsection{Stillinger's Laplacian for 
non-integer dimensional space}

For a function $\varphi=\varphi(r,\theta)$ of radial distance $r$ 
and related angle $\theta$ measured relative 
to an axis passing through the origin, 
the scalar Laplacian in a non-integer dimensional space
proposed by Stillinger \cite{Stillinger} is
\be \label{NI-1}
^{St}\Delta^{D} = \frac{1}{r^{D-1}} \frac{\partial}{\partial r} \left( r^{D-1} \,\frac{\partial}{\partial r} \right) +
\frac{1}{r^2 \, \sin^{D-2} \theta} 
\frac{\partial}{\partial \theta} 
\left(  \sin^{D-2} \theta 
\frac{\partial}{\partial \theta} \right) ,
\ee
where $D$ is the dimension of space ($0<D <3$),
and the variables $r \ge 0$, $0\le \theta \le \pi$.
Note that
$(\, ^{St}\Delta^{D} )^2 \ne \, ^{St}\Delta^{2D}$.
If the function depends on radial distance $r$ 
only ($\varphi=\varphi(r)$), then
\be \label{NI-R}
^{St}\Delta^{D} \varphi (r) = \frac{1}{r^{D-1}} \frac{\partial}{\partial r} \left( r^{D-1} \,\frac{\partial \varphi (r)}{\partial r} \right) =
\frac{\partial^2 \varphi (r)}{\partial r^2} +
\frac{D-1}{r} \, \frac{\partial \varphi (r)}{\partial r} .
\ee
It is easy to see that 
the Stillinger's form of Laplacian $\, ^{St}\Delta^{D}$
for radial scalar functions $\varphi ({\bf r})=\varphi (r)$
coincides with the scalar Laplacian 
$\, ^S\Delta^{D}_r$ defined by (\ref{S-Delta-D}), i.e.,
\be \label{NI-R-2}
^{St}\Delta^{D} \varphi (r) = \, ^S\Delta^{D} \varphi (r) .
\ee
The Stillinger's Laplacian can be applied 
for scalar fields only. It cannot be used 
to describe vector fields ${\bf v}=v_r(r) \, {\bf e}_r$ 
because this Laplacian for $D=3$ is not equal to 
the usual vector Laplacian for $\mathbb{R}^3$,
\be
^{St}\Delta^{3} {\bf v}(r) \ne \, \Delta {\bf v}(r) =
\Bigl( 
\frac{\partial^2 v_r}{\partial r^2} + \frac{2}{r} \, 
\frac{\partial v_r}{\partial r} - \frac{2}{r^2} \, v_r
\Bigr) \, {\bf e}_r  .
\ee
The gradient, divergence, curl operator and vector Laplacian are not considered by Stillinger in paper \cite{Stillinger}.

%%%%%%%%%%%%%%%%%%%%%%%%%%%%%%%%%%%%%%%%%%%%%%%%%%%%%%%%%%%%

\subsection{Differential operators for $d \ne D-1$}

Let us consider a ball region $B_D$ in the fractal fluids
with the boundary $S_d= \partial B_D$ with dimensions
\be
\operatorname{dim} (B_D)=D, \quad 
\operatorname{dim} (S_d) = d .
\ee 
Equations (\ref{Div-D}) - (\ref{V-Delta-DC}) 
define differential operators for spaces
with non-integer dimension $D$ and 
boundary dimensions $d=D-1$.
In general, the dimension $D$ of the region of a fractal fluid 
and the dimension $d$ of boundary of this region
are not related by the equation $d = D-1$, i.e.,
\be
\operatorname{dim} (\partial B_D) \ne D-1 . 
\ee

Using the integration in non-integer dimensional space
and the correspondent Gauss's theorem, we can define 
%\cite{CNSNS2014}
the divergence for the case $d \ne D-1$ by the equation
\be \label{Div-Dd}
\operatorname{Div}^{D,d}_{r} {\bf v} =\frac{\pi^{(d+1-D)/2} \, \Gamma (D/2) }{\Gamma ((d+1)/2)}
\left( \frac{1}{r^{D-1-d}}
\frac{\partial v_r(r)}{\partial r} + \frac{d}{r^{D-d}} \, v_r(r) 
\right) .
\ee
For $d=D-1$, we get (\ref{Div-D}).
We can define the parameter
%%% dimension along the radial direction by
\be \label{ar}
\alpha_r = D-d ,
\ee
that can be interpreted as a dimension of fractal fluid  
along the radial direction ${\bf e}_r$. Using (\ref{ar}), 
equation (\ref{Div-Dd}) can be rewritten in the form
\be \label{Div-Dd2}
\operatorname{Div}^{D,d}_{r} {\bf v} = 
\pi^{(1-\alpha_r)/2} \, 
\frac{\Gamma ((d+\alpha_r)/2) }{ \Gamma ((d+1)/2)}
\left( \frac{1}{r^{\alpha_r-1}}
\frac{\partial v_r(r)}{\partial r} + 
\frac{d}{r^{\alpha_r}} \, v_r(r) \right) .
\ee
This is divergence operator for non-integer boundary dimension
$d \ne D-1$.
For $\alpha_r=1$, we have $d=D-1$, and then equations 
(\ref{Div-Dd}), (\ref{Div-Dd2}) give (\ref{Div-D}).

The gradient for the scalar field $\varphi({\bf r})= \varphi(r)$ 
and the radial dimension $\alpha_r \ne 1$ is defined by
\be \label{Grad-Dd}
\operatorname{Grad}^{D,d}_{r} \varphi = 
\frac{\Gamma (\alpha_r/2)}{ \pi^{\alpha_r/2} \, r^{\alpha_r-1}} \,
\frac{\partial \varphi(r)}{\partial r} \, {\bf e}_r .
\ee
For $\alpha_r=1$, equation (\ref{Grad-Dd}) gives (\ref{Grad-D}).

Using the operators (\ref{Grad-Dd}) and (\ref{Div-Dd})
for the fields $\varphi=\varphi (r)$ and 
${\bf v}=v(r) \, {\bf e}_r$,
we can get the scalar and vector Laplace operators 
for the case $d \ne D-1$ by
\be
^S\Delta^{D,d}_r \varphi= 
\operatorname{Div}^{D,d}_r 
\operatorname{Grad}^{D,d}_{r} \varphi , \quad
^V\Delta^{D,d}_r {\bf v} = 
\operatorname{Grad}^{D,d}_r 
\operatorname{Div}^{D,d}_{r} {\bf v} .
\ee

The scalar Laplacian for $d \ne D-1$ 
for the field $\varphi=\varphi (r)$ is
\be \label{S-Delta-Dd}
^S\Delta^{D,d}_r \varphi= A(d,\alpha_r)
\left(  \frac{1}{r^{2 \alpha_r-2}} \, 
\frac{\partial^2 \varphi}{\partial r^2} + 
\frac{d+1-\alpha_r}{r^{2\alpha_r-1}} \, 
\frac{\partial \varphi}{\partial r} \right) ,
\ee
where
\be \label{Ada}
A(d,\alpha_r) = 
\frac{\Gamma ((d+\alpha_r)/2) \, \Gamma (\alpha_r/2)}{ 
\pi^{\alpha_r-1/2} \, \Gamma ((d+1)/2)} .
\ee

The vector Laplacian for $d \ne D-1$  
for the field ${\bf v}=v(r) \, {\bf e}_r$ is
\be \label{V-Delta-Dd}
^V\Delta^{D,d}_r {\bf v} = A(d,\alpha_r) 
\left( \frac{1}{r^{2 \alpha_r-2}} \, 
\frac{\partial^2 v_r}{\partial r^2} 
+ \frac{d+1-\alpha_r}{r^{2\alpha_r-1}} \, 
\frac{\partial  v_r}{\partial r} 
- \frac{d \, \alpha_r}{r^{2\alpha_r}} \,  v_r
\right) \, {\bf e}_r.
\ee

The vector differential operators (\ref{Grad-Dd}), 
(\ref{Div-Dd}), (\ref{S-Delta-Dd}) and (\ref{V-Delta-Dd})
allow us to describe complex fractal fluids with 
the boundary dimensions $d \ne D-1$
by continuum models with non-integer dimensional spaces.

%%%%%%%%%%%%%%%%%%%%%%%%%%%%%%%%%%%%%%%%%%%%%%%%%%%%
%%%%%%%%%%%%%%%%%%%%%%%%%%%%%%%%%%%%%%%%%%%%%%%%%%%%
%%%%%%%%%%%%%%%%%%%%%%%%%%%%%%%%%%%%%%%%%%%%%%%%%%%%
\section{Navier-Stokes equations in non-integer 
dimensional space for fractal fluid}

A motion of an incompressible viscous fractal fluid
in the framework of continuum model with 
non-integer dimensional space is described by the equations
\be \label{nse-1}
\operatorname{Div}^D_r {\bf v} =0 ,
\ee
\be \label{NSE}
\frac{d {\bf v}}{dt} = {\bf f}
- \frac{1}{\rho} \operatorname{Grad}^D_r p + 
\nu \, ^V\Delta^D_r {\bf v} , 
\ee
where ${\bf f}$ is the vector field of mass forces,
$\nu$ is the kinematic viscosity is the ratio of the dynamic viscosity $\mu$ to the density of the fluid $\rho$, and
$d/dt$ is the material derivative
\be \label{nse-2}
\frac{d {\bf v}}{dt} = \frac{\partial {\bf v}}{\partial t} +
\Bigl({\bf v},\operatorname{Grad}^D_r {\bf v}\Bigr) .
\ee
In equations (\ref{nse-1})-(\ref{nse-2}) 
the gradient $\operatorname{Grad}^{D}_r$, 
the divergence $\operatorname{Div}^{D}_{r}$,
and the vector Laplacian $^V\Delta^{D}_r$
are defined by equations
(\ref{Grad-D}), (\ref{Div-D}), (\ref{V-Delta-D})
for spherical symmetry.
For cylindrical symmetry,
these operators are defined by equations
(\ref{Grad-DC}), (\ref{Div-DC}), (\ref{V-Delta-DC}).

If the dimension $D$ of the region of a fractal fluid
and the dimension $d$ of boundary of this region
are not related by the relation $d = D-1$, i.e.,
$\alpha_r = D-d \ne 1$, then we should use the equations
\be \label{nse-1d}
\operatorname{Div}^{D,d}_r {\bf v} =0 ,
\ee
\be \label{NSEd}
\frac{d {\bf v}}{dt} = {\bf f}
- \frac{1}{\rho} \operatorname{Grad}^{D,d}_r p + 
\nu \, ^V\Delta^{D,d}_r {\bf v} , 
\ee
\be \label{nse-2d}
\frac{d {\bf v}}{dt} = \frac{\partial {\bf v}}{\partial t} +
\Bigl({\bf v},\operatorname{Grad}^{D,d}_r {\bf v}\Bigr) ,
\ee
where the gradient $\operatorname{Grad}^{D,d}_r$, 
the divergence $\operatorname{Div}^{D,d}_{r}$,
and the vector Laplacian $^V\Delta^{D,d}_r$
are defined by equations
(\ref{Grad-Dd}), (\ref{Div-Dd2}), (\ref{V-Delta-Dd}).

Equations (\ref{NSE}) and (\ref{NSEd}) 
can be called the Navier-Stokes equations
for non-integer dimensional space.

It is convenient to work in the dimensionless space variables
$x/R_0 \to x$, $y/R_0 \to x$, $z/R_0 \to x$, $r/R_0 \to r$,
that yields dimensionless integration and dimensionless 
differentiation in non-integer dimensional space.
Here $R_0$ is the characteristic size of a 
fractal fluid, which is always finite. 
For example, $R_0$ can be the minimal scale of self-similarity 
of a considered fractal fluid. 
Then the density is properly scaled such that
the mass $Q$ of fractal fluid and the fields ${\bf v}$, $p$, $f$
have correct physical dimensions.

The Navier-Stokes equations (\ref{NSE}) and (\ref{NSEd}) 
describe dynamics of fractal fluids in the framework of
continuum models with non-integer dimensional spaces.
These equations allow us to describe the isotropic fractal fluid 
only when the presence of spherical or cylindrical symmetry.

Equations (\ref{NSE}) and (\ref{NSEd}) can be used, when
the fields $p$, ${\bf v}$, ${\bf v}$ have the form
$p  =p(r)$ and ${\bf v}=v_r(r) \, {\bf e}_r$,
${\bf f})=f_r(r) \, {\bf e}_r$
does not depend on the angles.

As an example of application of the 
Navier-Stokes equations (\ref{NSE}) and (\ref{NSEd}),
we consider a steady flow of fractal fluid in a pipe, 
and fractal fluid discharge in the next sections.
In the next sections, we derive the Poiseuille equation 
for fractal fluids from the Navier-Stokes equations (\ref{NSE}) and (\ref{NSEd}). 
Equations can be used for any other problems of
hydrostatics and hydrodynamics of fractal fluids
within the case of spherical and cylindrical symmetries

To consider anisotropic fractal fluids,
and problems without spherical and cylindrical symmetries,
we cannot use the Navier-Stokes equations 
(\ref{NSE}) and (\ref{NSEd}).
In this case, we should apply a product measure approach
\cite{JMP2014}.
The product measure approach to describe fractal
properties of space-time has been considered in \cite{Svozil}.
The product measure approach for the fractional spaces
has been suggested in \cite{PRE2005,JPCS2005},
where fractional phase space is considered with its 
interpretation as a non-integer (fractional) dimensional space.
For non-integer dimensional spaces 
the product measure approach is suggested in \cite{PS2004},
where each orthogonal coordinates has own dimension. 
The product measure approach for 
the fractional-integral continuum models
has been considered in \cite{MOS-4,MOS-4b,MOS-4c,MOS-5}.

%%%%%%%%%%%%%%%%%%%%%%%%%%%%%%%%%%%%%%%%%%%%%%%%
\section{Steady flow of fractal fluid in a pipe} 

In this section, we derive the Poiseuille equation 
for fractal fluids from the Navier-Stokes equations (\ref{NSE}). 
Let us consider a simple problem of motion 
of an incompressible viscous fractal fluid.
Using the continuum models with non-integer dimensional space,
we describe a steady flow of fractal fluid 
in a pipe with circular cross-section. 
We take the axis of the pipe as the $X$-axis. 
The velocity for laminar of fractal fluid
is along the $X$-axis at all points, 
and is a function of $r$ only
\be 
{\bf v}={\bf v}(r)=v_x(r) \, {\bf e}_x .
\ee
The equation of continuity is satisfied identically. 
The components of the Navier-Stokes equation 
for $Y$-axis and $Z$-axis 
give that the pressure is constant over 
the cross-section of the pipe. 

We shall solve the equation for a pipe 
with circular cross-section. 
Taking the origin at the center of the circle 
we can use cylindrical symmetry
${\bf v}={\bf v}(r)=v_x(r) \, {\bf e}_x$. 
Using the Navier-Stokes equations (\ref{NSE}), we have 
\be \label{NSE-X}
^S\Delta^D_r v_x(r)= \frac{1}{ \mu} \,\frac{dp}{dx} ,
\ee
where $\mu= \rho \, \nu$, and $dp/dx$ is a constant. 
The pressure gradient $dp/dx$ may be written $ - \Delta p /l$, 
where $\Delta p$ is the pressure difference between 
the ends of the pipe and $l$ is its length. 

%%%%%%%%%%%%%%%%%%%%%%%%%%%%%%%%%

Using the scalar Laplacian 
for non-integer dimensional space
the Navier-Stokes equation (\ref{NSE-X}) takes the form
\be \label{Cyl-1}
\frac{\partial^2 v_x(r)}{\partial r^2} + \frac{D-2}{r} \, 
\frac{\partial v_x}{\partial r} - \frac{1}{ \mu} \,\frac{dp}{dx} = 0 .
\ee
For $1<D<3$ and $0<D<1$, 
the general solution of (\ref{Cyl-1}) is
\be \label{vxD}
v_x(r) =C_1 \, r^{3-D} + C_2 + \frac{1}{2 \, (D-1) \, \mu} 
\frac{dp}{dx} \, r^2 \quad (0<D<3, \quad D \ne 1).
\ee
For $D=3$, we have
\be
v_x(r) =C_1 \, \ln(r) + C_2 + \frac{1}{4 \, \mu} 
\frac{dp}{dx} \, r^2 .
\ee
For $D=1$, we get the general solution
\be
v_x(r) =C_1 \, r^2 + C_2 + \frac{1}{4 \, \mu} 
\frac{dp}{dx} \, r^2 \, \left(2 \, \ln(r) - 1 \right) .
\ee
It should be noted that dimensions $D=1$ of the fractal fluid 
do not correspond to the distribution of particles 
along the line. 
The fractal media with $D=1$
describe a distribution of fluid particles 
in 3-dimensional space 
such that the mass dimension of the distribution 
is equal to $D=1$.

%%%%%%%%%%%%%%%%%%%%

Let us determine a flow of fractal fluid in a pipe 
of annular cross-section with 
the internal radius $R_1$ and external radius $R_2$. 
The constants $C_1$ and $C_2$ 
in the general solution (\ref{vxD}) 
are determined from the boundary conditions 
\be \label{vR1vR2}
v_x(R_1) = v_x(R_2) = 0 .
\ee
Using (\ref{vxD}), these conditions have the form
\be
C_1 \, R_1^{3-D} + C_2 + \frac{1}{2 \, (D-1) \, \mu} 
\frac{dp}{dx} \, R_1^2 =0 ,
\ee
\be
C_1 \, R_2^{3-D} + C_2 + \frac{1}{2 \, (D-1) \, \mu} 
\frac{dp}{dx} \, R_2^2 =0 .
\ee
Then the constants are
\be \label{C1}
C_1 = \frac{1}{2 \, (D-1) \, \mu} \frac{dp}{dx} \, 
\frac{R_2^2- R_1^2}{R_1^{3-D}-R_2^{3-D}} ,
\ee
\be \label{C2}
C_2 = \frac{1}{2 \, (D-1) \, \mu} \frac{dp}{dx} 
\, \frac{ R_1^{3-D}\, R_2^2- R_2^{3-D} \, R_1^2}{R_1^{3-D}-R_2^{3-D}} .
\ee
Substitution of (\ref{C1}) and (\ref{C2}) into (\ref{vxD}) gives
\be \label{vxr}
v_x(r) = \frac{1}{2 \, (D-1) \, \mu} \, \frac{dp}{dx}
\left( \frac{ R_2^2- R_1^2}{R_1^{3-D}-R_2^{3-D}} \, r^{3-D} + 
\frac{R_1^{3-D}\, R_2^2- R_2^{3-D} \, R_1^2}{R_1^{3-D}-R_2^{3-D}} +  r^2 \right) ,
\ee
where $0<D<1$ and $1<D <3$.

Using the variables
\[ x=\frac{R_2}{R_1} , \quad  y=\frac{r}{R_1} , \]
equation (\ref{vxr}) can be represented in the form
\be \label{vxr2}
v(x,y) = \frac{R^2_1}{2 \, (D-1) \, \mu} \, \frac{dp}{dx}
\left( - \frac{1-x^2}{1-x^{3-D}} \, y^{3-D} + 
\frac{x^2 - x^{3-D}}{1-x^{3-D}} +  y^2 \right) ,
\ee
To demonstrate some properties of the velocity
$v_x(r)$ defined by (\ref{vxr}), 
we can visualize the function (\ref{vxr2}), 
for $x \in [1;100]$, $y \in [1;100]$ 
and different values of dimensions 
$D=2.9$, $D=2.7$, $D=2.0$, $D=1.1$.
The plots of function (\ref{vxr2}) are presented by Figures 1-4, 
where $\mu=1$, $R_1=1$ and ${dp}/{dx}=-1$.

%%%%%%%%%%%%%%%%%%%%%%%%%%%%%%%%%%%%%%%%%%%%%%%%%%%%%%%%%%%%%%%%
%%%%%%%%%%%%%%%%%%%%%%%%%%%%%%%%%%%%%%%%%%%%%%%%%%%%%%%%%%%%%%%%
%%%%%%%%%%%%%%%%%%%%%%%%%%%%%%%%%%%%%%%%%%%%%%%%%%%%%%%%%%%%%%%%

%%% ----------------------- PLOTS 1 -------------------------

%%%\newpage
%%%\setcounter{figure}{1}
%%%%%%%%%%%%%%%%%%%%%%%%%%%%%%%%%%%%%%%%%%%%%%%%%%%%%%%%%%%%%%
\begin{figure}[H]
\begin{minipage}[h]{0.47\linewidth}
\resizebox{11cm}{!}{\includegraphics[angle=-90]{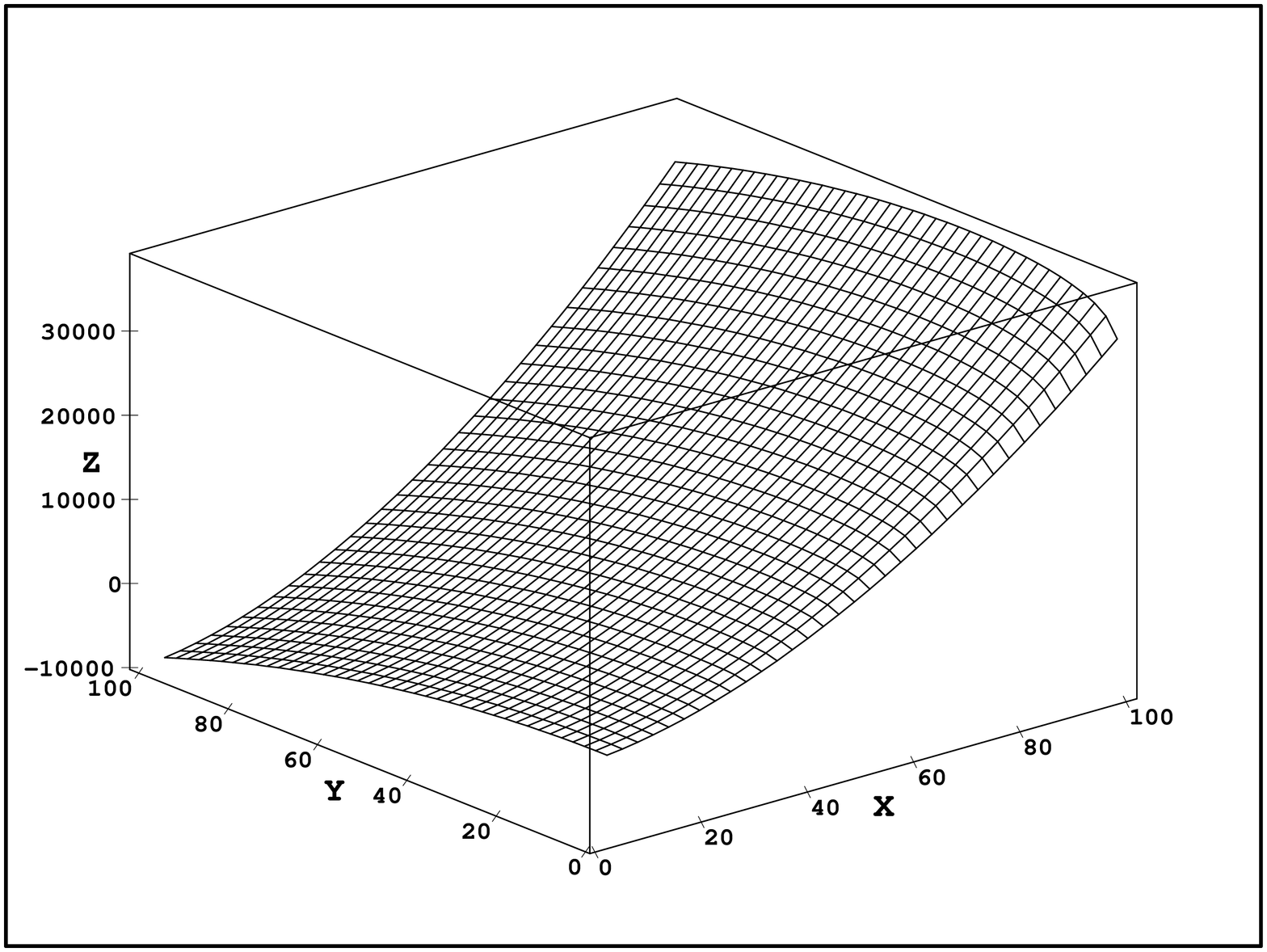}} 
\end{minipage}
\caption{Plot of the velocity function $z=v(x,y)$ 
defined by (\ref{vxr2}) for the ranges $x \in [1;100]$,  
$y \in [1;100]$, and $D=2.9$.} 
\label{Plot1}
\end{figure}

%%% ----------------------- PLOTS 2 -------------------------

\begin{figure}[H]
\begin{minipage}[h]{0.47\linewidth}
\resizebox{11cm}{!}{\includegraphics[angle=-90]{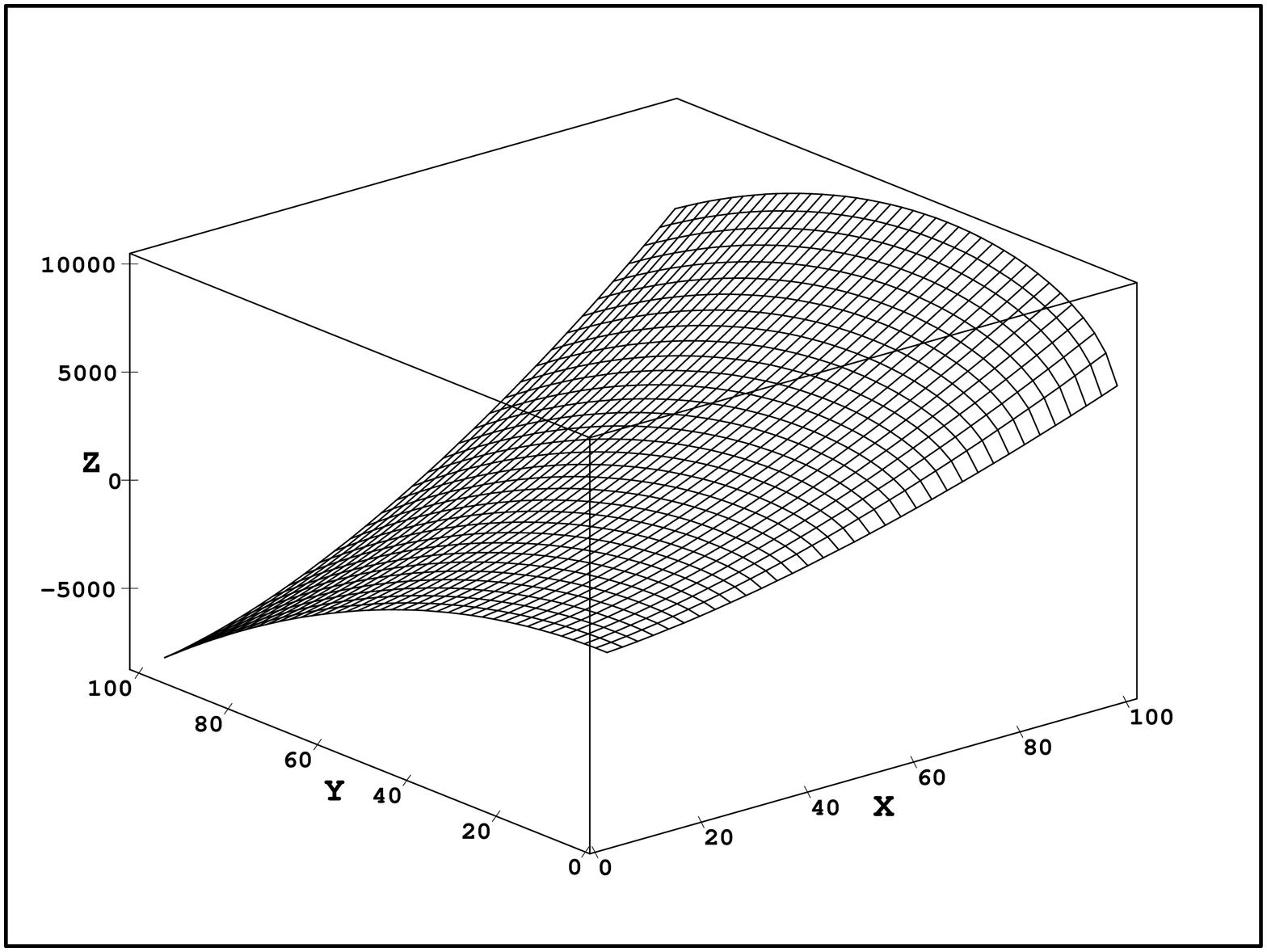}} 
\end{minipage}
\caption{Plot of the velocity function $z=v(x,y)$ 
defined by (\ref{vxr2}) for the ranges $x \in [1;100]$,  
$y \in [1;100]$, and $D=2.7$.} 
\label{Plot2}
\end{figure}

%%% ----------------------- PLOTS 3 -------------------------

\begin{figure}[H]
\begin{minipage}[h]{0.47\linewidth}
\resizebox{11cm}{!}{\includegraphics[angle=-90]{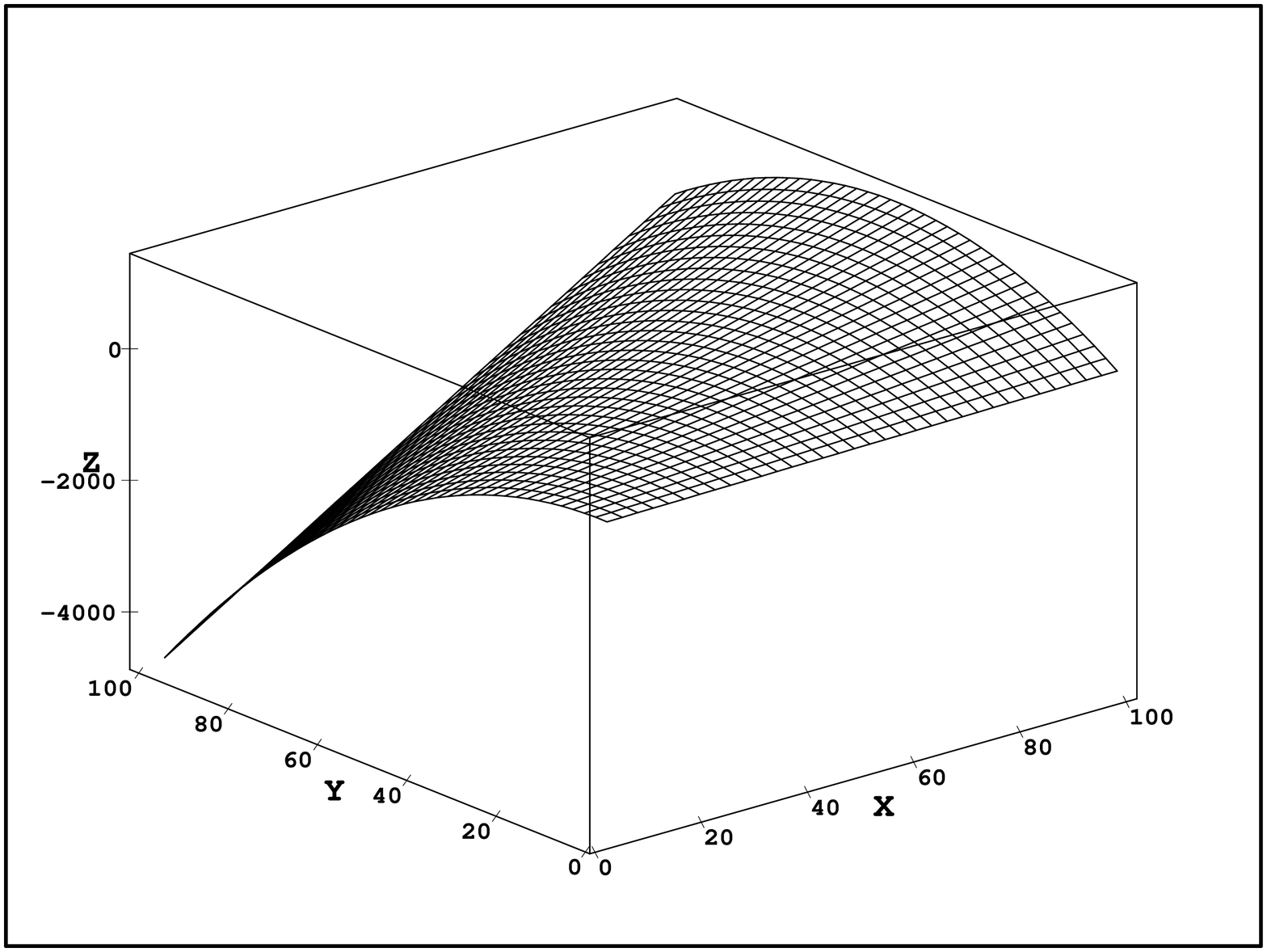}} 
\end{minipage}
\caption{Plot of the velocity function $z=v(x,y)$ 
defined by (\ref{vxr2}) for the ranges $x \in [1;100]$,  
$y \in [1;100]$, and $D=2.0$.} 
\label{Plot3}
\end{figure}

%%% ----------------------- PLOTS 4 -------------------------

\begin{figure}[H]
\begin{minipage}[h]{0.47\linewidth}
\resizebox{11cm}{!}{\includegraphics[angle=-90]{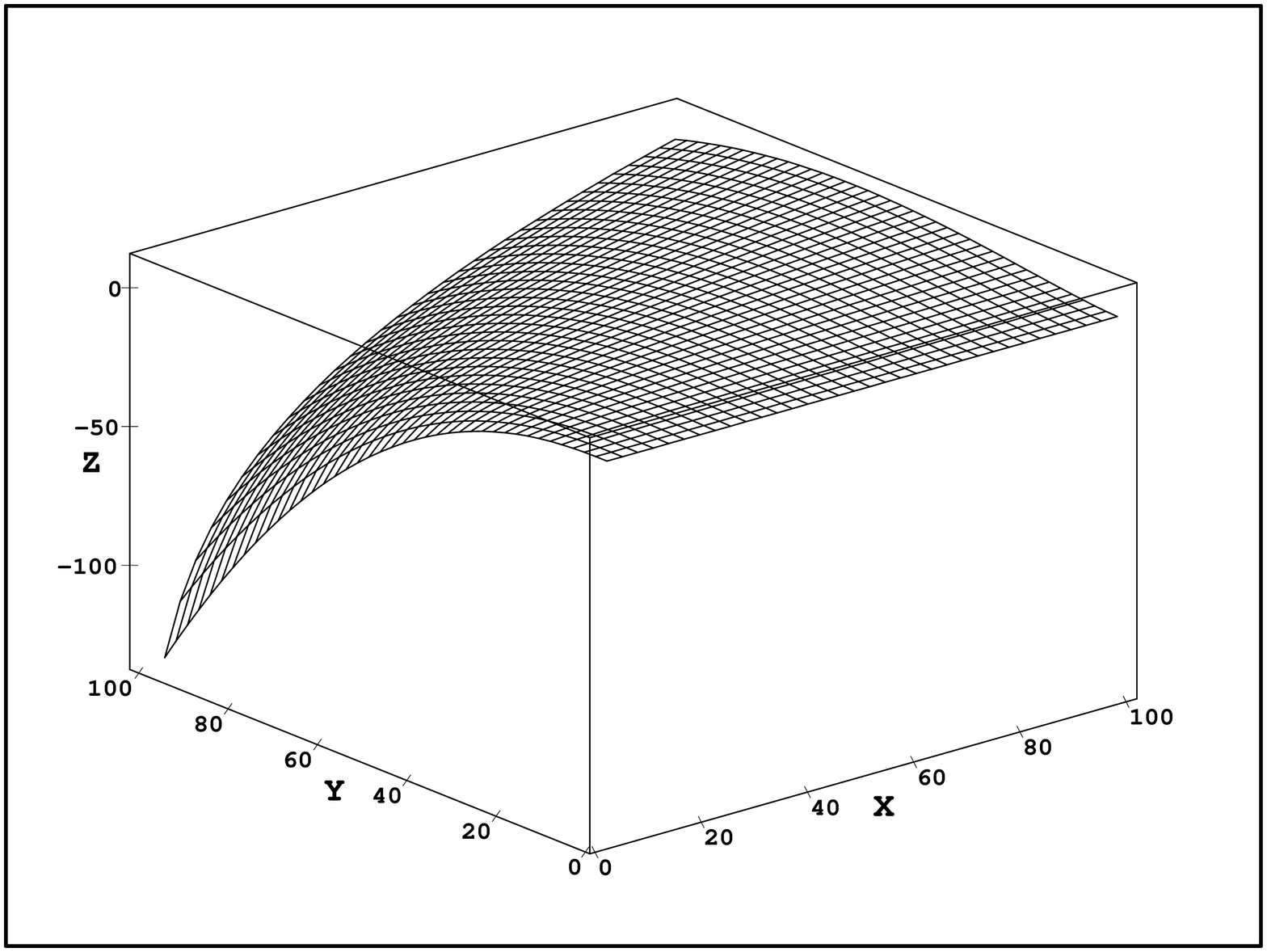}} 
\end{minipage}
\caption{Plot of the velocity function $z=v(x,y)$ 
defined by (\ref{vxr2}) for the ranges $x \in [1;100]$,  
$y \in [1;100]$, and $D=1.1$.} 
\label{Plot4}
\end{figure}

%%%%%%%%%%%%%%%%%%%%%%%%%%%%%%%%%%%%%%%%%%%%%%%%%%%%%%%%%%%%%%%%
%%%%%%%%%%%%%%%%%%%%%%%%%%%%%%%%%%%%%%%%%%%%%%%%%%%%%%%%%%%%%%%%
%%%%%%%%%%%%%%%%%%%%%%%%%%%%%%%%%%%%%%%%%%%%%%%%%%%%%%%%%%%%%%%%

The flow in a pipe of annular cross-section  
with the radius $R$, i.e. $R_1=0$ and $R_2=R$, we have 
\be \label{PE-FF}
v_x(r) = - \frac{1}{2 \, (D-1) \, \mu} \frac{dp}{dx} \, R^2
\left( \left(\frac{r}{R}\right)^{3-D} 
- \left(\frac{r}{R}\right)^2 \right) \quad
(0<D \le 3, \quad D \ne 1).
\ee
Equation (\ref{PE-FF}) can be called 
the Poiseuille equation for flow of fractal fluid. 
For the case of non-fractal fluid ($D=3$), 
equation (\ref{PE-FF}) gives the well-known Poiseuille equation
\be \label{PE-FF-D3}
v_x(r) = - \frac{1}{4 \, \mu} \frac{dp}{dx} \, R^2
\left( 1 - \left(\frac{r}{R}\right)^2 \right) .
\ee
Thus the velocity distribution 
across the pipe is parabolic for the non-fractal fluids. 
For the fractal fluids, we have 
non-integer power-law (\ref{PE-FF}). 

Note that suggested Poiseuille equation 
for fractal fluids, which are derived 
from the proposed Navier-Stokes equation 
for non-integer dimensional space 
can be used only to describe 
flow of fractal fluid in pipes.
To describe flow of fractal fluid between parallel planes, we should define new differential operators 
and the Navier-Stokes equations (\ref{NSE}) and (\ref{NSEd}), 
should be modified by using the product measure approach
\cite{Svozil,PS2004,PRE2005,JPCS2005}.
%%% \cite{Product2014}

%%%%%%%%%%%%%%%%%%%%%%%%%%%%%%%%%%%%%%%%%%%%%%%%%
\section{Fractal fluid with $\alpha_r \ne 1$}

Let us derive the Poiseuille equation for fractal fluids 
from the Navier-Stokes equations (\ref{NSEd}). 
The Navier-Stokes equations for 
fractal fluid with $\alpha_r=D-d \ne 1$ has the form
\be \label{Eq-Dda}
A(d_x,\alpha_r) \left(  \frac{1}{r^{2 \alpha_r-2}} \, 
\frac{\partial^2 v_x(r)}{\partial r^2} + 
\frac{d_x +1-\alpha_r}{r^{2\alpha_r-1}} \, 
\frac{\partial v_x(r)}{\partial r} \right) -
\frac{1}{ \mu} \,\frac{dp}{dx} = 0 ,
\ee
where $A(d_x,\alpha_r)$ is defined by (\ref{Ada}),
$d_x=d- \alpha_x$, and
$\alpha_x$ is dimension along the $X$-axis.
Using $v_x(r)$ as an effective scalar field
$\varphi_{eff}(r)=v_x(r)$, we can apply equations
(\ref{Grad-Dd}), (\ref{Div-Dd2}) and (\ref{S-Delta-Dd})
where $D \to D_x=D- \alpha_x$ and $d \to d_x=d- \alpha_x$
to get (\ref{Eq-Dda}).
Equation (\ref{Eq-Dda}) with $\alpha_r=\alpha_x=1$ 
gives (\ref{Cyl-1}).

For $1<D<3$ and $0<D<1$, 
the general solution of (\ref{Eq-Dda}) is
\be \label{vxDda}
v_x(r) =C_1 \, r^{\alpha_r-d_x} + C_2 + 
\frac{1}{2 \, (d_x+ \alpha_r) \, \alpha_r \, A(d_x,\alpha_r) \, \mu} 
\frac{dp}{dx} \, r^{2 \alpha_r} 
\quad (0<D<3, \quad D \ne 1) .
\ee
For $\alpha_r=\alpha_x=1$ equation (\ref{vxDda}) gives (\ref{vxD}).
The constants $C_1$ and $C_2$ 
in the general solution (\ref{vxDda}) 
are determined by the boundary conditions 
\be \label{vR1vR2b}
v_x(R_1) = v_x(R_2) = 0 .
\ee
These conditions give the equations
\be 
C_1 \, R_1^{\alpha_r-d_x} + C_2 + 
\frac{1}{2 \, (d_x+ \alpha_r) \, \alpha_r \, A(d_x,\alpha_r) \, \mu} 
\frac{dp}{dx} \, R_1^{2 \alpha_r} =0 ,
\ee
\be 
C_1 \, R_2^{\alpha_r-d_x} + C_2 + 
\frac{1}{2 \, (d_x+ \alpha_r) \, \alpha_r \, A(d_x,\alpha_r) \, \mu} 
\frac{dp}{dx} \, R_2^{2 \alpha_r} =0 .
\ee
Then the coefficients are
\be \label{C1da}
C_1 = - \frac{1}{2 \, (d_x+ \alpha_r) \, 
\alpha_r \, A(d_x,\alpha_r) \, \mu} \frac{dp}{dx} 
\frac{ R_1^{2\alpha_r} - R_2^{2\alpha_r} }{R_1^{\alpha_r-d_x} - R_2^{\alpha_r-d_x}} ,
\ee
\be \label{C2da}
C_3 = - \frac{1}{2 \, (d_x+ \alpha_r) \, 
\alpha_r \, A(d_x,\alpha_r) \, \mu} \frac{dp}{dx} 
\frac{ R_2^{2\alpha_r} R_1^{\alpha_r-d_x} - R_1^{2\alpha_r} R_2^{\alpha_r-d_x} }{ R_1^{\alpha_r-d_x} - R_2^{\alpha_r-d_x}} ,
\ee
Substitution of (\ref{C1da}) and (\ref{C2da}) 
into (\ref{vxDda}) gives
\[
v_x(r) = - \frac{1}{2 \, (d_x+ \alpha_r) \, 
\alpha_r \, A(d_x,\alpha_r) \, \mu} \frac{dp}{dx} \Bigl(
\frac{ R_1^{2\alpha_r} - R_2^{2\alpha_r} }{R_1^{\alpha_r-d_x} - R_2^{\alpha_r-d_x}} \, r^{\alpha_r-d_x} + \]
\be \label{vxDda2}
+ \frac{ R_2^{2\alpha_r} R_1^{\alpha_r-d_x} - R_1^{2\alpha_r} R_2^{\alpha_r-d_x} }{ R_1^{\alpha_r-d_x} - R_2^{\alpha_r-d_x}} 
- r^{2 \alpha_r} \Bigr) 
\ee
for $0<D<3$, where $D \ne 1$.

If $R_1=0$ and $R_2=R$, then equation (\ref{vxDda2}) has the form
\be \label{vxDda3}
v_x(r) = - \frac{1}{2 \, (d_x+ \alpha_r) \, 
\alpha_r \, A(d_x,\alpha_r) \, \mu} \frac{dp}{dx} \,
R^{2\alpha_r}  \, \left(
\left(\frac{r}{R}\right)^{\alpha_r-d_x} -
\left( \frac{r}{R}\right)^{2 \alpha_r} \right) .
\ee
For $\alpha_r=\alpha_x=1$ 
equation (\ref{vxDda3}) gives (\ref{PE-FF}).
The cases $\alpha_r <1$ and/or $\alpha_x <1$ 
corresponds to fractal fluid.

We can assume that $\alpha_x >1$ can be used 
to describe fractal turbulent flow in pipe.
This assumption is based on the fact
that trajectories of the fluid particles 
are fractal curve, then $\alpha_x>1$
(for example, the Koch curve with 
$\alpha_x=\ln(4) / \ln(3) \approx 1.262$).

%%%%%%%%%%%%%%%%%%%%%%%%%%%%%%%%%%%%%%%%%%%%%%%%%
\section{Fractal fluid discharge}

In general, fractal fluids cannot be considered 
as a fluid on fractal.
Real fractal fluids have a characteristic smallest length scale
such as the radius, $R_0$, of a particle 
(for example, an atom or molecule).
In real fluids the fractal structure cannot be observed on all
scales but only those for which $R >R_0$, 
where $R_0$ is the characteristic scale of the particles.
The concept of non-integer mass dimension of fractal fluid
is based on the idea of how the mass 
of a fluid region scales with the region size,
if we assume unchanged density.
For many cases, we can write the asymptotic form for the relation between
the mass $M_D(W)$ of a ball region $W$ of fluid,
and the radius $R$ containing this mass as follows:
\be \label{1-1-MR} 
M_D(W)=M_0 \left( \frac{R}{R_0} \right)^D  \ee
for $R/R_0 \gg 1$.
The constant $M_0$ depends on how the spheres of radius $R_0$ 
are packed.
The parameter $D$, which is interpreted as a dimension, 
does not depend on the shape of the region $W$,
or on whether the packing of spheres of radius $R_0$ is 
close packing,
a random packing or a porous packing with a uniform distribution of holes.
The non-integer mass dimension $D$ of fractal fluid 
is a measure of how the fluid fills 
the integer $n$-dimensional Euclidean space it occupies.
Note that the fact that a fluid is random or contains cavities
does not necessarily imply that the fluid is fractal.

Using the non-integer dimensional space approach,
we can calculate the mass of fractal homogeneous fluids. 
Scaling law (\ref{1-1-MR}) is obtained naturally 
in the framework of this approach.
We can use the integration in a non-integer dimensional space 
\cite{Stillinger} that is described by the equation
\be \label{NI-2}
\int_{R^D} d^D {\bf r} \, \varphi ({\bf r}) =
\frac{2 \pi^{(D-1)/2}}{\Gamma((D-1)/2)}
\int^{\infty}_0 dr \, r^{D-1} \,
\int^{\pi}_0 d \theta \, \sin^{D-2}\theta \, \varphi (r,\theta) ,
\ee
where $d^D {\bf r}$ represent the volume element
in the non-integer dimensional space.
Using (\ref{NI-2}) with $\varphi (r,\theta)=1$, and
\be \int^{\pi}_0 d \theta \, sin^{D-2}\theta = 
\frac{\pi^{1/2}\, \Gamma (D/2-1)}{\Gamma(D/2)} , \ee
we get the volume of $D$-dimensional ball with radius $R$ in the form
\be
V_D = \frac{\pi^{D/2}}{\Gamma(D/2+1)} \, R^{D} .
\ee
The mass of fluid in $W$ is described by the integral 
\be \label{1-MW3} 
M_D(W) = \int_{W} \rho({\bf r}) \, d^D {\bf r} ,
\ee
where ${\bf r}$ is dimensionless vector variable. 
For a ball with radius $R$ and constant density 
$\rho({\bf r})=\rho =\operatorname{const}$, we get 
\be \label{M-D}
M_D(W) = \rho \, V_D =
\frac{\pi^{D/2} \, \rho}{\Gamma(D/2+1)} \, R^{D} .
\ee
This equation define the mass of the 
fractal homogeneous ball region of fluid with volume $V_D$. 
For $D=3$, equation (\ref{M-D}) gives the well-known 
equation for mass of non-fractal ball region
$M_3 =(4 \rho \pi /3) \, R^3$ because 
$\Gamma(3/2) =\sqrt{\pi}/2$ and $\Gamma(z+1)=z \, \Gamma(z)$. 

%%%%%%%%%%%%%%%%%%%%%%%%%%%%%%%%%%%%%%%%%%%%%%%

Let us determine the mass $Q$ of fluid passing per unit time 
through any cross-section of the pipe (called the discharge). 
Not all pipe volume is occupied by fractal fluid.
There are areas unoccupied by particles of fractal fluid.
In continuum model of fractal fluids, 
we take into account this fact by using the integration 
in space with non-integer dimension
\be \label{Q-Def}
Q= \rho \, \frac{2 \, \pi^{d/2}}{\Gamma(d/2)} 
\int^R_0 v_x(r) \, r^{d-1} \, dr ,
\ee
where $d=D-1$ is non-integer dimension of 
the cross-section, $\rho$ is a constant density, 
and $v_x(r)$ is defined by equation (\ref{PE-FF}).
Substitution of (\ref{PE-FF}) into (\ref{Q-Def}) gives
\be \label{QR}
Q= - \frac{\rho \, \pi^{(D-1)/2}}{2 \, (D+1) \, \Gamma((D-1)/2)\, \mu} 
\frac{dp}{dx} \, R^{D+1} \quad (0<D \le 3 \quad D \ne 1).
\ee
Note that $\rho$ has physical dimension of mass 
(for example, kilogram).
The mass of fractal fluid is thus proportional 
to $(D+1)$-power of the radius of the pipe.

For $D=3$, equation (\ref{QR}) gives the well-known equation
\be \label{QRD3}
Q= - \frac{\rho \, \pi}{8\, \mu} \frac{dp}{dx} \, R^{4} .
\ee
The mass of non-fractal fluid is proportional to 
the fourth power of the radius of the pipe. 
The dependence of $Q$ on $dp/dx$ and $R$ given by formula 
(\ref{QRD3}) was established empirically by G. Hagen in 1839 and J. L. M. Poiseuille in 1840, and theoretically justified by G. G. Stokes in 1845.

Equation (\ref{QR}) can be rewritten in the form
\be
Q= - \frac{\rho \, \pi}{8\, \mu_{eff}} \frac{dp}{dx} \, R^{D+1} 
\ee
with the effective dynamic viscosity 
\be \label{meff}
\mu_{eff} = 
\frac{D+1}{4} \, \pi^{(3-D)/2} \ \Gamma((D-1)/2) \ \mu .
\ee
The mass (discharge) of fractal fluid is proportional to 
the non-integer power $(D+1)<4$ of the radius of the pipe,
and the dynamic viscosity is effectively changed. 

For $D=3$, equation (\ref{meff}) gives $\mu_{eff}=\mu$.
For $1<D<3$, we have $\mu_{eff}>\mu$.
If $0<D<1$, then $\mu_{eff} <0$.
The effective dynamic viscosity of 
fractal fluid with $D \in (1;3)$ 
increases with increasing a deviation of the dimension $D$ 
from three. 
We can see an interesting effect of 
a negative effective dynamic viscosity for fractal fluid 
with dimension $D \in (0;1)$. 
The strong fractality of the fluid, 
which is caused by small dimension, 
leads to an increased fluid flow 
compared with conventional medium.
This is probably due to an increase in freedom of particles motion for the fractal fluid similar to 3D Cantor dust.

The discharge function $Q=Q(R,D)$ defined by (\ref{QR}) 
for the different values of dimensions $0 < D < 3$ 
and the range $R\in[0,1]$ are present on Figures 5-8, 
where $\rho=1$, $\mu=1$, and ${dp}/{dx}=-1$.

%%%%%%%%%%%%%%%%%%%%%%%%%%%%%%%%%%%%%%%%%%%%%%%%%%%%%%%%%%%%%%%%
%%%%%%%%%%%%%%%%%%%%%%%%%%%%%%%%%%%%%%%%%%%%%%%%%%%%%%%%%%%%%%%%
%%%%%%%%%%%%%%%%%%%%%%%%%%%%%%%%%%%%%%%%%%%%%%%%%%%%%%%%%%%%%%%%

%%% ----------------------- PLOTS 5 -------------------------

%%%\newpage
%%%\setcounter{figure}{1}
%%%%%%%%%%%%%%%%%%%%%%%%%%%%%%%%%%%%%%%%%%%%%%%%%%%%%%%%%%%%%%
\begin{figure}[H]
\begin{minipage}[h]{0.47\linewidth}
\resizebox{11cm}{!}{\includegraphics[angle=-90]{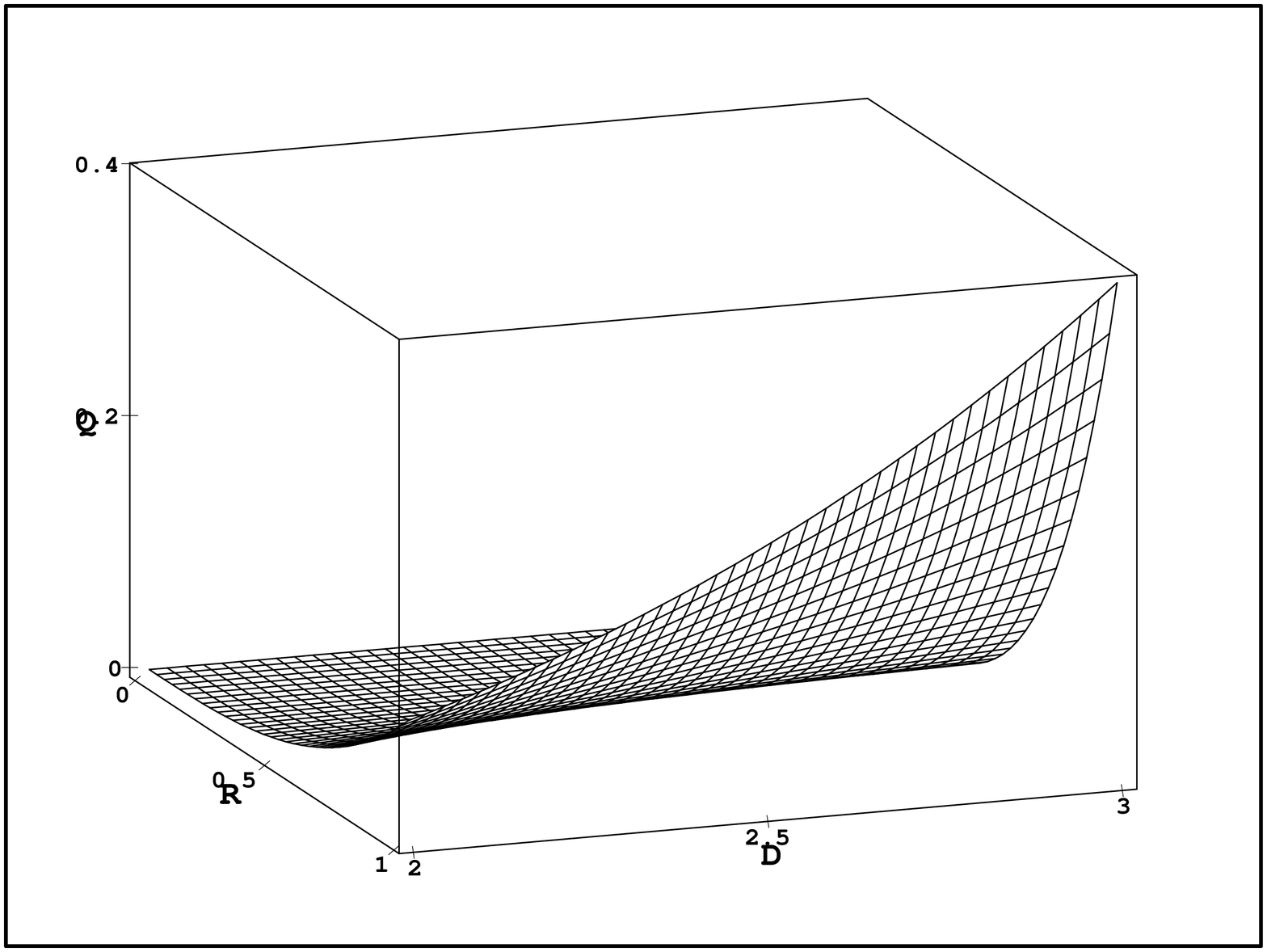}} 
\end{minipage}
\caption{Plot of the discharge function $Q=Q(R,D)$ defined by (\ref{QR}) for the range $R\in[0,1]$ and $D\in[2,3]$.} 
\label{Plot5}
\end{figure}

%%% ----------------------- PLOTS 6 -------------------------

\begin{figure}[H]
\begin{minipage}[h]{0.47\linewidth}
\resizebox{11cm}{!}{\includegraphics[angle=-90]{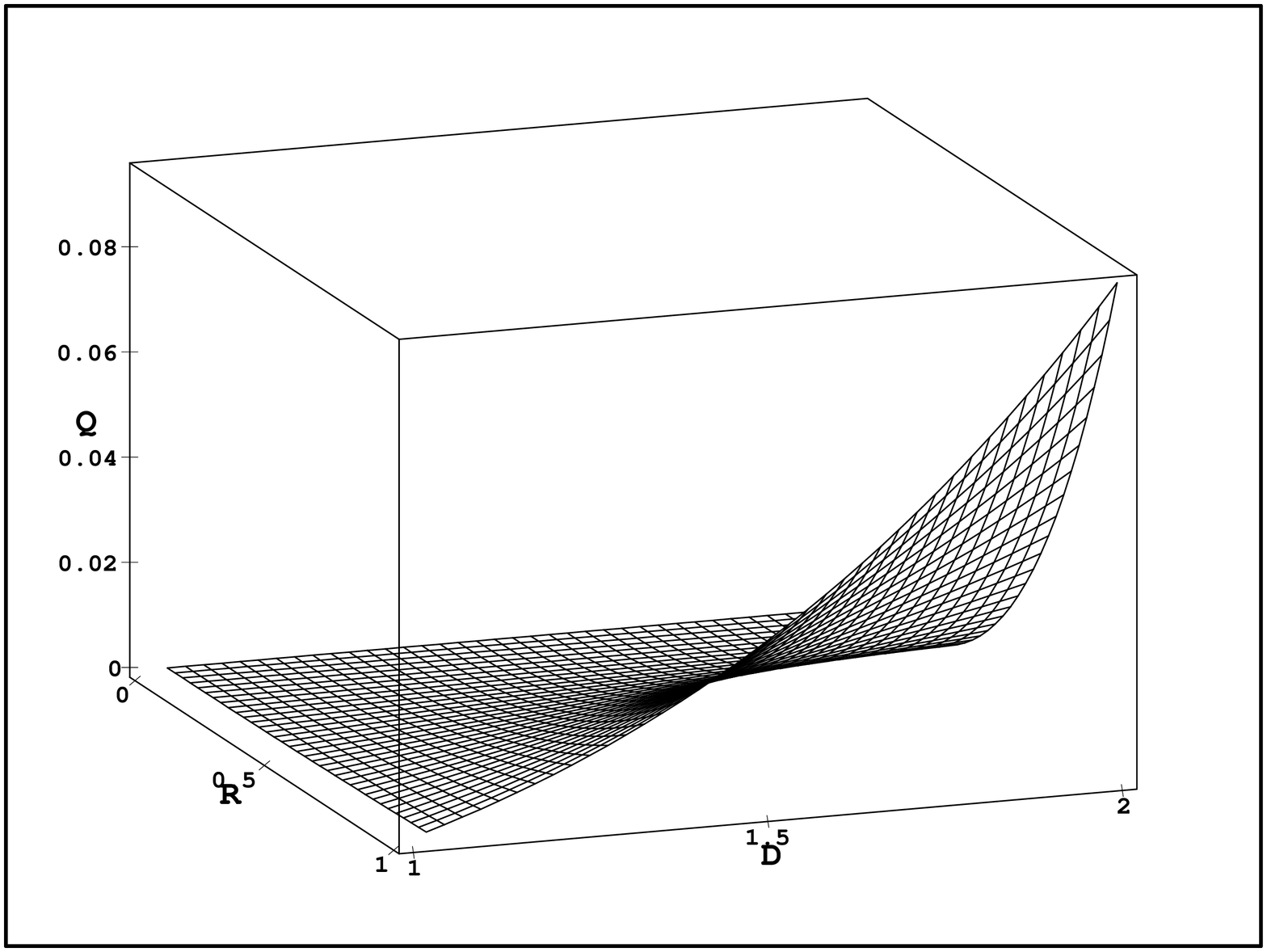}} 
\end{minipage}
\caption{Plot of the discharge function $Q=Q(R,D)$ defined by (\ref{QR}) for the range $R\in[0,1]$ and $D\in[1,2]$.} 
\label{Plot6}
\end{figure}

%%% ----------------------- PLOTS 7 -------------------------

\begin{figure}[H]
\begin{minipage}[h]{0.47\linewidth}
\resizebox{11cm}{!}{\includegraphics[angle=-90]{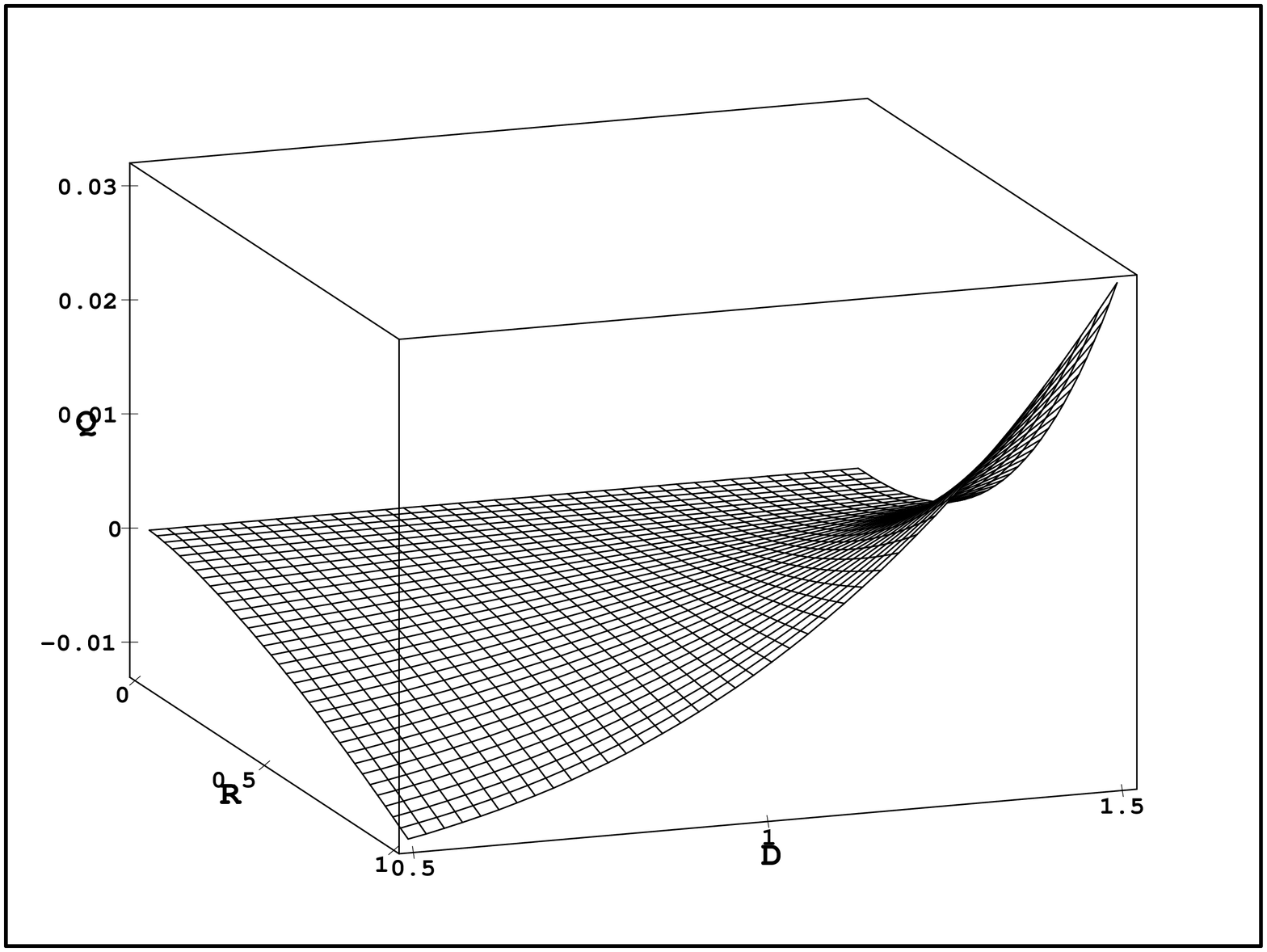}} 
\end{minipage}
\caption{Plot of the discharge function $Q=Q(R,D)$ defined by (\ref{QR}) for the range $R\in[0,1]$ and $D\in[0.5,1.5]$.} 
\label{Plot7}
\end{figure}

%%% ----------------------- PLOTS 8 -------------------------

\begin{figure}[H]
\begin{minipage}[h]{0.47\linewidth}
\resizebox{11cm}{!}{\includegraphics[angle=-90]{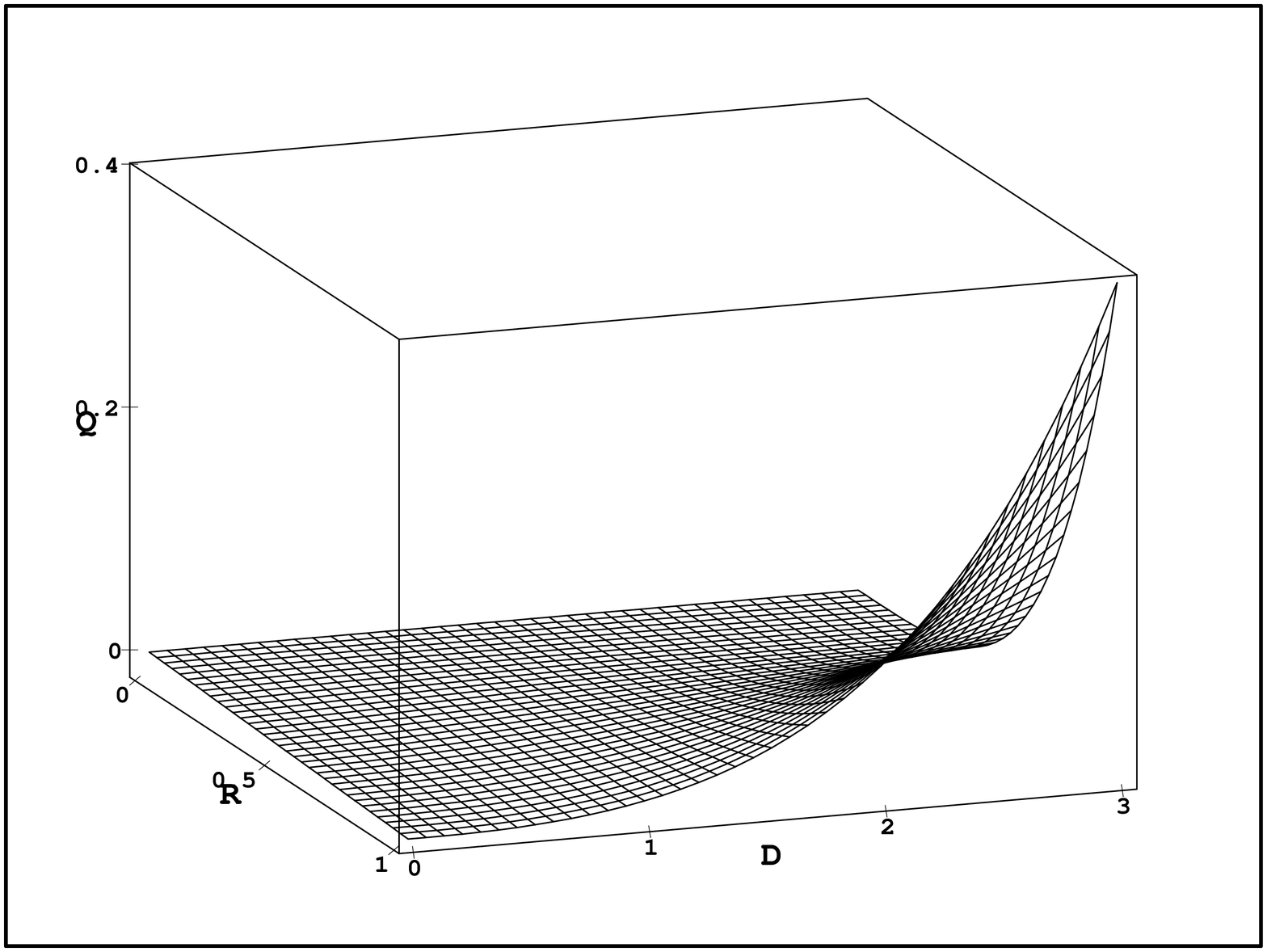}} 
\end{minipage}
\caption{Plot of the discharge function $Q=Q(R,D)$ defined by (\ref{QR}) for the range $R\in[0,1]$ and $D\in[0,3]$.} 
\label{Plot8}
\end{figure}

%%%%%%%%%%%%%%%%%%%%%%%%%%%%%%%%%%%%%%%%%%%%%%%%%%%%%%%%%%%%%%%
%%%%%%%%%%%%%%%%%%%%%%%%%%%%%%%%%%%%%%%%%%%%%%%%%%%%%%%%%%%%%%%
%%%%%%%%%%%%%%%%%%%%%%%%%%%%%%%%%%%%%%%%%%%%%%%%%%%%%%%%%%%%%%%

\section{Conclusion}

In this paper, we propose a generalization of the Navier-Stokes 
equations to describe fractal fluids in the framework of
continuum models with non-integer dimensional spaces. 
These equations contain generalized 
differential vector operators for
non-integer dimensional space \cite{CNSNS2015,JMP2014}.
As an example of application of the suggested
Navier-Stokes equations for fractal fluids, 
we consider a Poiseuille flow of 
an incompressible viscous fractal fluid in the pipe. 
The solution for steady flow of 
fractal fluid in a pipe and corresponding 
fractal fluid discharge have been derived.

In this paper fractal fluid is described as a continuum 
in non-integer dimensional space. 
We assume that suggested continuum models with non-integer
dimensional spaces and the correspondent
Navier-Stokes equations for fractal fluids
may be important for fractal theory of different type of media.

%%% The concept of fractal fluid

As the main object for application of the proposed 
continuum models is a two-component medium, 
where distribution of one component  (gas, liquid, solid) 
into another component (fluid, gas or empty space) 
can be characterized by non-integer mass or 
"particle" dimension.
This non-integer dimensional component can be considered 
as a fractal fluid. 
One of the possible experimental methods 
for determining the presence of 
fractal properties of the two-component medium 
may be to use labels with radioactive isotopes 
for particles of component that is assumed a fractal.

%%% Liquid + Empty Space

A basic idealized model of fractal fluid 
is a liquid distributed in empty space $\mathbb{R}^3$
with non-integer mass dimension $D<3$.
In some sense the fractal fluid is considered 
as a liquid analog of fractal porous solid material.
%%% Liquid + Gas
Fractal fluid can also be viewed as a two-phase medium 
consisting of a liquid and a discharged gas instead 
of empty space,
where the liquid is characterized by fractal mass dimension.

%%% Emulsion, Solution = liquid + liquid

As an object of study, we also can consider 
an emulsion, when both the dispersed and the continuous phase 
are liquids, and the dispersed phase is fractally distributed 
in continuous phase.
An emulsion that is a mixture of two immiscible liquids,
one of which (the dispersed phase) is fractally dispersed 
in the other (the continuous phase). 
In this case the dispersed phase can be described 
as a fractal fluid by suggested continuum models 
with non-integer dimensional space.
The proposed models can be used for
a solution that is a homogeneous mixture 
composed of only one (liquid) phase,
where one phase has a fractal dimension. 
We can consider a fractal distribution of a solute 
dissolved in a non-fractal solvent,
then the solute is considered as a fractal fluid.
The solvent that is fractally homogeneously mixed with solute 
can be considered as a fractal homogeneous fluid.
The homogeneity property of the fractal fluid means 
that two regions $W_1$ and $W_2$ 
with the equal volumes $V_n(W_1)=V_n(W_2)$ 
have equal number of particles $N_D(W_1)=N_D(W_2)$. 
In other words the fractal fluid is called homogeneous 
if the power law $N_D(W) \sim R^{D}$ 
(or $M_D(W) \sim R^{D}$) does not depend on 
the translation of the region $W$. 

%%% Suspension = solid + liquid

We can consider a fractal distribution of small solid particles 
in the suspension. In this case, we have 
an internal phase (solid) that is fractally distributed through 
the external phase (fluid) by mechanical agitation. 

An object of investigations can be a liquid mixed 
with a solid particles, 
where the distribution of these particles in space
can be characterized by non-integer mass dimensions,
which can be caused by a power law distribution 
of particles by size or mass.

%%%%%%%%%%%%%  blood

As a complex medium which may exhibit 
fractal properties can be considered the blood
that is composed of
proteins, glucose, mineral ions, hormones, 
carbon dioxide, blood cells 
and other particles suspended in water. 
We assume that the blood as a multi-phase medium
can have attributes of a fractal distribution for 
some blood components including bacteria, 
viruses and medicinal substances getting into the blood.

%%%%%%%%%%%%%%%%%%%%%%%%%%%%%%%%%%

We assume that the suggested approach to describe 
fractal fluids by continuum models with non-integer
dimensional spaces may be important for
fractal theory of blood flow in cardiovascular system, 
dynamics of fractal media in hydrologic modeling 
\cite{FLU-1,FLU-2,FLU-3,FLU-4} and 
it allows to develop the fractal dynamics 
of multi-phase media \cite{Soo,Nigmatulin}.

%%%%%%%%%%%%%%%%%%%%%%%%%%%%%%%%%%%%%%%%%%%%%%%%%%%%%%%%%%%%%%%
%%%%%%%%%%%%%%%%%%%%%%%%%%%%%%%%%%%%%%%%%%%%%%%%%%%%%%%%%%%%%%%
%%%%%%%%%%%%%%%%%%%%%%%%%%%%%%%%%%%%%%%%%%%%%%%%%%%%%%%%%%%%%%%

%%%%%%%%%%%%%%%%%%%%%%%%%%%%%%%%%%%%%%%%%%%%%%%%%%%%%%%%%%%%%%%
%%%%%%%%%%%%%%%%%%%%%%%%%%%%%%%%%%%%%%%%%%%%%%%%%%%%%%%%%%%%%%%
%%%%%%%%%%%%%%%%%%%%%%%%%%%%%%%%%%%%%%%%%%%%%%%%%%%%%%%%%%%%%%%

\end{document}